\documentclass[manuscript]{aastex}

\slugcomment{2015 November 26 version}

\shorttitle{Formation Mechanisms of HC$_{5}$N in TMC-1}
\shortauthors{K. Taniguchi et al.}

\begin{document}


\title{Implication of Formation Mechanisms of HC$_{5}$N in TMC-1 \\
       as Studied by $^{13}$C Isotopic Fractionation}


\author{Kotomi Taniguchi\altaffilmark{1,2}, Hiroyuki Ozeki\altaffilmark{1}, Masao Saito\altaffilmark{2,3}, Nami Sakai\altaffilmark{4,5}, Fumitaka Nakamura\altaffilmark{2,6}, Seiji Kameno\altaffilmark{2,7}, Shuro Takano\altaffilmark{2,3,8}, and Satoshi Yamamoto\altaffilmark{4}}

\email{kotomi.taniguchi@nao.ac.jp}


\altaffiltext{1}{Department of Environmental Science, Faculty of Science, Toho University, Miyama, Funabashi, Chiba 274-8510, Japan}
\altaffiltext{2}{Department of Astronomical Science, School of Physical Science, SOKENDAI (The Graduate University for Advanced Studies), Osawa, Mitaka, Tokyo 181-8588, Japan} 
\altaffiltext{3}{Nobeyama Radio Observatory, National Astronomical Observatory of Japan, Minamimaki, Minamisaku, Nagano 384-1305, Japan}
\altaffiltext{4}{Department of Physics, Faculty of Science, The University of Tokyo, Hongo, Bunkyo, Tokyo 113-0033, Japan}
\altaffiltext{5}{The Institute of Physical and Chemical Research (RIKEN), Wako, Saitama 351-0198, Japan}
\altaffiltext{6}{National Astronomical Observatory of Japan, Osawa, Mitaka, Tokyo 181-8588, Japan}
\altaffiltext{7}{Joint ALMA Observatory, Alonso de Cordova 3107 Vitacura, Santiago 763 0355, Chile}
\altaffiltext{8}{Department of Physics, General Studies, College of Engineering, Nihon University, Koriyama, Fukushima 963-8642, Japan}


\begin{abstract}
We observed the {\it J} = $9-8$ and $16-15$ rotational transitions of the normal species and five $^{13}$C isotopologues of HC$_{5}$N to study its formation mechanisms toward the cyanopolyyne peak in Taurus Molecular Cloud-1, with the 45-m radio telescope of Nobeyama Radio Observatory.
We detected the five $^{13}$C isotopologues with high signal-to-noise ratios between 12 and 20, as well as the normal species.
The abundance ratios of the five $^{13}$C isotopologues of HC$_{5}$N are found to be $1.00:0.97:1.03:1.05:1.16$ ($\pm 0.19$) (1$\sigma$) for $[$H$^{13}$CCCCCN$]: [$HC$^{13}$CCCCN$]: [$HCC$^{13}$CCCN$]: [$HCCC$^{13}$CCN$]: [$HCCCC$^{13}$CN$]$.
We do not find any significant differences among the five $^{13}$C isotopologues.
The averaged [HC$_{5}$N]/[$^{13}$C isotopologues] abundance ratio is determined to be $94 \pm 6$ (1$\sigma$), which is slightly higher than the local interstellar elemental $^{12}$C/$^{13}$C ratio of $60-70$. 
Possible formation pathways are discussed on the basis of these results.

\end{abstract}





\section{Introduction}\label{intoro}

More than 180 molecules have been detected in the interstellar medium and circumstellar shells of evolved stars, and approximately 40\% of them are classified into carbon-chain molecules.
It is therefore of fundamental importance for astrochemistry to study their formation processes. 
However, these molecules are so reactive due to unsaturated chemical bonds and/or unpaired electrons that laboratory measurements of their reaction rates are not routine experiments.
Although formation mechanisms of the carbon-chain molecules have mainly been studied by chemical model calculations \citep[e.g.][]{mce13,wak15}, it is still difficult to reproduce molecular abundances derived by observations, and to determine formation mechanisms of carbon-chain molecules, because of uncertain rate coefficients and of poor knowledge of elementary reactions involving carbon-chain molecules.

Another method to study major formation pathways of molecules is based mainly on observations.
Recent developments of radio astronomical instruments allow us to detect low-abundance species, including rare isotopologues, with a reasonable observation time.
Formation mechanisms of some representative carbon-chain molecules have been investigated by observing their $^{13}$C isotopic fractionation, such as HC$_{3}$N \citep{tak98}, CCH \citep{sak10}, CCS \citep{sak07}, C$_{3}$S, and C$_{4}$H \citep{sak13}, toward the cyanopolyyne peak in Taurus Molecular Cloud-1 (TMC-1 CP; $d = 140$ pc), and {\it {cyclic}}-C$_{3}$H$_{2}$ \citep{yos15} toward the low-mass star-forming region L1527.
TMC-1 CP is a representative cold dark cloud where various carbon-chain molecules are abundant.
For instance, \citet{kai04} carried out spectral line survey observations in the $8.8-50$ GHz region toward TMC-1 CP with the 45-m radio telescope of Nobeyama Radio Observatory (NRO), and demonstrated that this source is rich in carbon-chain molecules.   

Cyanoacetylene, HC$_{3}$N, is a very abundant carbon-chain molecule in TMC-1 CP.
\citet{tak98} observed three $^{13}$C isotopologues of HC$_{3}$N toward TMC-1 CP, and the relative abundance of HCC$^{13}$CN is significantly higher than that of the other isotopologues: $[$H$^{13}$CCCN$]: [$HC$^{13}$CCN$]: [$HCC$^{13}$CN$]$ is $1.0 : 1.0 : 1.4$.
These results imply that HC$_{3}$N is mainly formed by the reaction between a hydrocarbon molecule with two equivalent carbon atoms and a molecule with one carbon atom, whose $^{12}$C/$^{13}$C ratios are different from each other.
According to {\it {ab initio}} calculations, the reaction of C$_{2}$H$_{2}$ + CN is exothermic, and has no significant energy barrier \citep{woo97,fuk98}.
The laboratory experiments show that the rate coefficient for this reaction is sufficiently large ($k \geq 4\times10 ^{-10}$ cm$^{3}$ molecule$^{-1}$ s$^{-1}$) at low temperatures \citep{sim93}.
Considering these results, \citet{tak98} suggested that HC$_{3}$N is formed by the reaction between C$_{2}$H$_{2}$ and CN.

It can naturally be predicted that HC$_{5}$N is formed by the similar mechanism of HC$_{3}$N, the reaction of C$_{4}$H$_{2}$ + CN.
\citet{fuk98} conducted {\it {ab initio}} calculations on the reactions of C$_{2n}$H$_{2}$ + CN ($n = 1-4$), which form HC$_{2n+1}$N, and found that these reactions are all exothermic and have no energy barriers.
In addition, \citet{sek96} measured the rate constant for the reaction C$_{4}$H$_{2}$ + CN to be ($4.2 \pm 0.2$)$\times$10$^{-10}$ cm$^{3}$ molecule$^{-1}$ s$^{-1}$ at the room temperature.
These results suggest that HC$_{5}$N can be formed by the reaction of C$_{4}$H$_{2}$ + CN.
If so, we would be able to observe different abundances of the five $^{13}$C isotopologues of HC$_{5}$N. 

Observations of the $^{13}$C isotopologues of HC$_{5}$N were carried out toward TMC-1 CP by \citet{tak90} . 
Although they detected the five $^{13}$C isotopologues of HC$_{5}$N, they could not determine well their relative abundances due to low signal-to-noise ratios ($3-6$).
\citet{tak98} also reported the spectra of the two $^{13}$C isotopologues, HC$^{13}$CCCCN and HCCCC$^{13}$CN, with slightly higher signal-to-noise ratios ($\approx 8$).
However, they could not discuss its main formation pathway, because of the lack of data for the other three $^{13}$C isotopologues.

In the present paper, we observed the five $^{13}$C isotopologues of HC$_{5}$N toward TMC-1 CP with the NRO 45-m telescope with high enough signal-to-noise ratios to constrain mechanisms responsible for the formation of HC$_{5}$N.

\section{Observations}\label{obs}  


We carried out observations of the normal species and the five $^{13}$C isotopologues of HC$_{5}$N with the NRO 45-m radio telescope in 2014 March, April (2013-2014 season), December, and 2015 January (2014-2015 season).
The observed position was ($\alpha_{2000}$, $\delta_{2000}$) = (04$^{\rm h}$41$^{\rm m}$42\fs49, 25\arcdeg41\arcmin27\farcs0) for TMC-1 CP. 
The Z45 receiver \citep{nak15} and the high electron mobility transistor (HEMT) amplifier receiver (H22), both of which can obtain dual-polarization data simultaneously, were used for the observations of the {\it J} = $16-15$ and {\it J} = $9-8$ transitions at 42 GHz and 23 GHz, respectively. 
The exact frequencies of the observed lines are given in Table \ref{tab1}.
The system temperatures of the Z45 and the H22 receivers were from 100 to 130 K, and from 90 to 110 K, respectively.
The beam sizes and the main beam efficiencies ($\eta_{B}$) were 37" and 0.72 for 42 GHz \citep{nak15}, and 72" and 0.8 for 23 GHz. 
The telescope pointing was checked every 1.5 hours by observing the SiO maser line ({\it J} = 1-0) from NML Tau and GL 5134, and the pointing error was less than 3". 
We used the SAM45 FX-type digital correlator in frequency settings whose bandwidths and resolutions are 125 MHz and 30.52 kHz, respectively, for observations at 42 GHz, and 63 MHz and 15.26 kHz, respectively, for those at 23 GHz. 
These frequency resolutions correspond to the velocity resolution of about 0.2 km s$^{-1}$.
We employed the position-switching mode, where the off-source position was set to be +30' away in the right ascension.
The smoothed bandpass calibration method \citep{yam12} was adopted in the analysis, which allows us to greatly improve the signal-to-noise ratios, compared with the standard position-switch observations.
In fact, the scan pattern of this method was a set of 20 seconds and 5 seconds for on-source and off-source positions, respectively.
We applied 60 and 32 channel-smoothing for the off-source spectra at 23 GHz and 42 GHz, respectively.




\section{Results and Analysis}

\subsection{Results}

\begin{figure}
\plotone{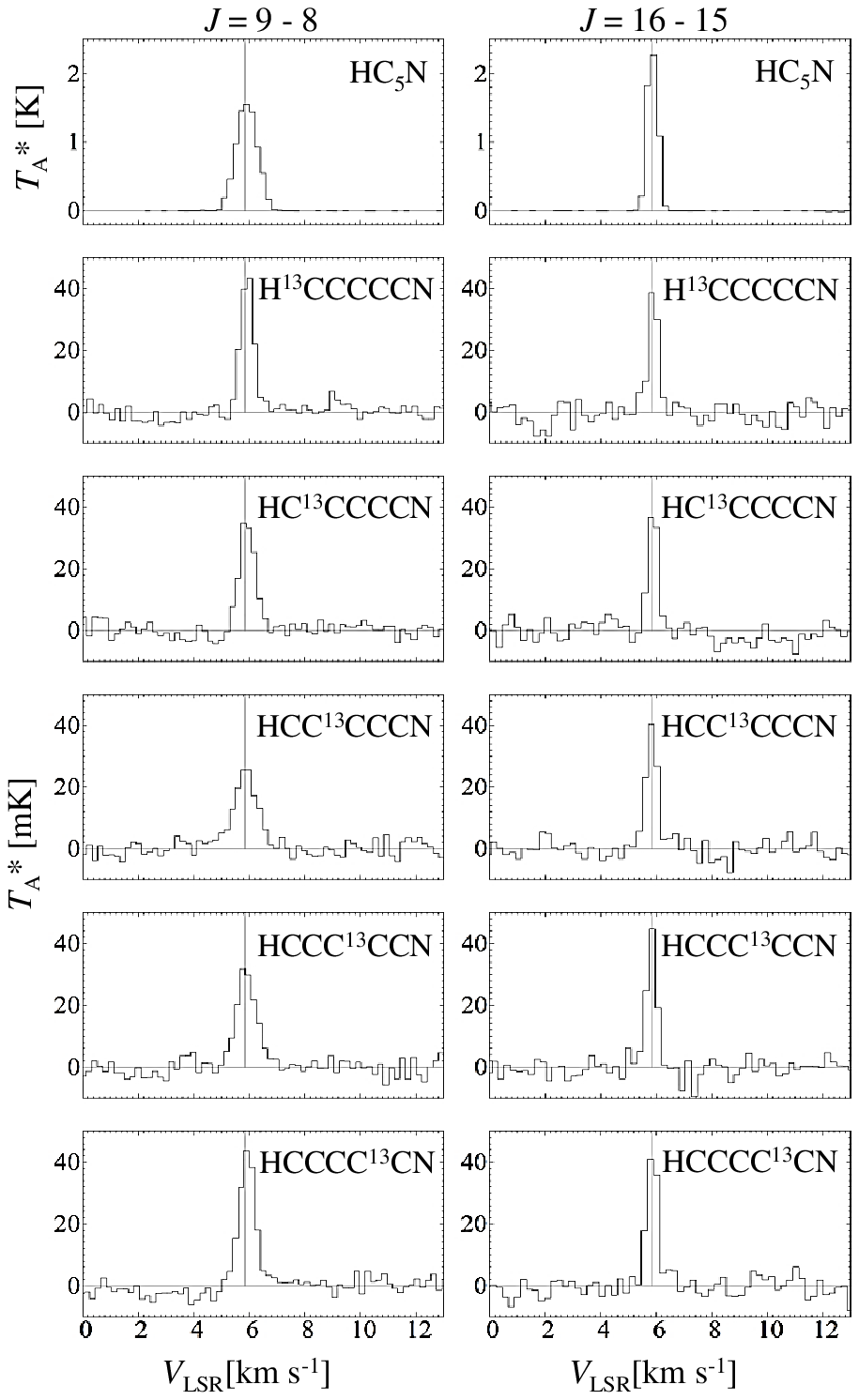}
\caption{Spectra of HC$_{5}$N and its $^{13}$C isotopologues of the {\it J} = $9-8$ and $16-15$ rotational transitions toward TMC-1 CP. HC$^{13}$CCCCN and HCCCC$^{13}$CN, HCC$^{13}$CCCN and HCCC$^{13}$CCN have been observed in the same IF bands. The vertical lines show {\it V}$_{\mathrm {LSR}}$ = 5.85 km s$^{-1}$. } 
\label{fig1}
\end{figure}

The spectra of the five $^{13}$C isotopologues of HC$_{5}$N were taken with signal-to-noise ratios of $12-20$, as shown in Figures \ref{fig1}.
We also obtained high quality data for the normal species.
Since the spectra show a symmetric single peak structure, we fitted the spectra with a Gaussian profile, and obtained the spectral line parameters, as summarized in Table \ref{tab1}.
The line widths ($\Delta v$ [km s$^{-1}$]) in the 23 GHz region ($\sim 0.8$ km s$^{-1}$) are broader than those in the 42 GHz region ($\sim 0.5$ km s$^{-1}$). 
Although the origin of their difference is unclear, the following possibility can be considered.
It is well known that the TMC-1 CP region has a complex velocity structure \citep{lan95,dic01}.
Since the telescope beam samples a wider area at 23 GHz than at 42 GHz, the line width could be larger at 23 GHz.
However, this difference does not affect our discussion on the relative abundance ratios among the five $^{13}$C isotopologues seriously, and hence, we do not discuss the line width further.
The values of {\it V}$_{{\mathrm {LSR}}}$ are consistent with one another, and are in good agreement with the {\it V}$_{{\mathrm {LSR}}}$ value reported for this source (5.85 km s$^{-1}$, \citet{kai04}).

The ratio of the integrated intensities ($\int T^{\ast}_{\mathrm A}dv$ [K km s$^{-1}$]) of the five $^{13}$C isotopologues is derived to be $1.00 : 0.95 : 0.97 : 1.01 : 1.16$ ($\pm 0.16$) (1$\sigma$) and $1.00 : 1.01 : 1.12 : 1.13 : 1.17$ ($\pm 0.19$) (1$\sigma$) for $[$H$^{13}$CCCCCN$]: [$HC$^{13}$CCCCN$]: [$HCC$^{13}$CCCN$]: [$HCCC$^{13}$CCN$]: [$HCCCC$^{13}$CN$]$ for the {\it J} = $9-8$ and {\it J} = $16-15$ lines, respectively. 
Although HCCCC$^{13}$CN is slightly brighter than the other species, there are no significant differences in intensities among the five $^{13}$C isotopologues.
The result contrasts with that for HC$_{3}$N, which shows clear differences in intensities of the three $^{13}$C isotopologues \citep{tak98}.

\subsection{Analysis}

The rotational population of HC$_{5}$N is not completely thermalized to the gas kinetic temperature, because the H$_{2}$ density of TMC-1 CP ($4 \times 10^{4}$ cm$^{-3}$ \citep{hir92}) is equal to  or lower than the critical densities of the HC$_{5}$N lines.
However, \citet{tak90} showed that the observed brightness temperatures of the HC$_{5}$N lines are well approximated by the rotation temperature of 6.5 K.
Then, we calculated the column densities of the normal species and the five $^{13}$C isotopologues of HC$_{5}$N using the local thermodynamic equilibrium (LTE) as shown the following formulae \citep{tak98}:

\begin{equation} \label{tau}
\tau = - {\mathrm {ln}} \left[1- \frac{T^{\ast}_{\rm A} }{f\eta_{\rm B} \left\{J(T_{\rm {ex}}) - J(T_{\rm {bg}}) \right\}} \right],  
\end{equation}
where
\begin{equation} \label{tem}
J(T) = \frac{h\nu}{k}\Bigl\{\exp\Bigl(\frac{h\nu}{kT}\Bigr) -1\Bigr\} ^{-1},
\end{equation}  
and
\begin{equation} \label{col}
N = \tau \frac{3h\Delta v}{8\pi ^3}\sqrt{\frac{\pi}{4\mathrm {ln}2}}Q\frac{1}{\mu ^2}\frac{1}{J_{\rm {lower}}+1}\exp\Bigl(\frac{E_{\rm {lower}}}{kT_{\rm {ex}}}\Bigr)\Bigl\{1-\exp\Bigl(-\frac{h\nu }{kT_{\rm {ex}}}\Bigr)\Bigr\} ^{-1}.
\end{equation} 
In equation (\ref{tau}), $T^{\ast}_{\rm A}$ denotes the antenna temperature, {\it f} the beam filling factor, $\eta_{\rm B}$ the main beam efficiency (see Section \ref{obs}), and $\tau$ the optical depth.
We used 0.8 and 1 for {\it f} in the 23 GHz and 42 GHz band data, respectively.
We employed these {\it f} values, because a size of the emitting region of carbon-chain molecules in TMC-1 CP is approximately 2.5' according to the mapping observations by \citet{hir92}.
$T_{\rm {ex}}$ is the excitation temperature, $T_{\rm {bg}}$ the cosmic microwave background temperature ($\approx 2.7$ K), and {\it J}({\it T}) in equation (\ref{tem}) the Planck function. 
We derived $T_{ex}$ and $\tau$ simultaneously from the observed intensities of the 23 GHz and 42 GHz lines by using the formula (1).
In equation (\ref{col}), {\it N} is the column density, $\Delta v$ the line width (FWHM), $Q$ the partition function, $\mu$ the permanent electric dipole moment of HC$_{5}$N ($4.33 \times 10^{-18}$ [esu cm]\citep{ale76}), and $E_{\rm {lower}}$ the energy of the lower rotational energy level. 
In the energy level calculations, we used the rotational constants of the normal species and the $^{13}$C isotopologues \citep{ale76,san05}. 

The calculated column densities and the $^{12}$C/$^{13}$C ratios are summarized in Table \ref{tab2}.
For the normal species, the excitation temperature and the column density are determined to be $6.5 \pm 0.2$ K and ($6.2 \pm 0.3$)$\times 10^{13}$ cm$^{-2}$, respectively, where the errors are estimated from the Gaussian fits, and represent the confidence level of one standard deviation.
The optical depths for the {\it J} = $9-8$ and $16-15$ lines are $1.084 \pm 0.014$ and $1.19 \pm 0.03$ (1$\sigma$), respectively. 
The obtained excitation temperature and column density of the normal species are consistent with the previous result ($6.5 \pm 0.2$ K and ($6.3 \pm 0.6$)$\times 10^{13}$ cm$^{-2}$) \citep{tak90} within the error ranges.

We calculated the column densities of the five $^{13}$C isotopologues using equation (\ref{col}) and the excitation temperature derived for the normal species (6.5 K).
The calculated column densities of the normal species and the five $^{13}$C isotopologues, and the $^{12}$C/$^{13}$C ratios are summarized in Table \ref{tab2}. 
For the five $^{13}$C isotopologues, the column densities were derived to be ($6.4 \pm 0.9$)$\times 10^{11}$, ($6.2 \pm 0.8$)$\times 10^{11}$, ($6.6 \pm 0.8$)$\times 10^{11}$, ($6.7 \pm 0.9$)$\times 10^{11}$, and ($7.4 \pm 0.9$)$\times 10^{11}$ cm$^{-2}$, for H$^{13}$CCCCCN, HC$^{13}$CCCCN, HCC$^{13}$CCCN, HCCC$^{13}$CCN, and HCCCC$^{13}$CN, respectively.
The abundance ratio of the $^{13}$C isotopologues of HC$_{5}$N is then $1.00:0.97:1.03:1.05:1.16$ ($\pm 0.19$) (1$\sigma$) for $[$H$^{13}$CCCCCN$]: [$HC$^{13}$CCCCN$]: [$HCC$^{13}$CCCN$]: [$HCCC$^{13}$CCN$]: [$HCCCC$^{13}$CN$]$. 
This result agrees with those obtained from integrated intensities.
The averaged $^{12}$C/$^{13}$C ratio of HC$_{5}$N is $94 \pm 6$ (1$\sigma$), which is slightly higher than the elemental ratio of $60-70$  in the local interstellar medium \citep[e.g.,][]{lan90,lan93,sav02,mil05}.
The intensities are also affected by the calibration errors (main beam efficiency and filling factor).
However, we obtained the data of the normal species and the five $^{13}$C isotopologues simultaneously.
Then, these calibration errors do not affect the $^{12}$C/$^{13}$C ratios.
We discuss these ratios in section \ref{dil}.

We investigated the sensitivities of the assumed  $T_{\rm {ex}}$ values on the column densities and the $^{12}$C/$^{13}$C ratios.
We calculated the column densities of the normal species and the five $^{13}$C isotopologues using $T_{\rm {ex}}$ of 5.9 K and 7.1 K, which are the lowest and highest values in the 3$\sigma$ error range of the excitation temperature.
We derived the column densities of the normal species to be ($7.0 \pm 0.5$)$\times 10^{13}$ cm$^{-2}$ and ($4.7 \pm 0.2$)$\times 10^{13}$ cm$^{-2}$ for 5.9 K and 7.1 K, respectively.
The column densities of the $^{13}$C isotopologues were not significantly affected by the change of the excitation temperature.
The $^{12}$C/$^{13}$C ratios averaged for the five isotopologues are $125 \pm 6$ (1$\sigma$) and $89 \pm 5$ (1$\sigma$) for 5.9 K and 7.1 K, respectively.
Although the  $^{12}$C/$^{13}$C ratios slightly depend on the excitation temperature, the $^{12}$C/$^{13}$C ratio of HC$_{5}$N is still higher than that of the elemental ratio in the local interstellar medium.
 
We also evaluated the ratio by using the excitation temperature derived for each $^{13}$C isotopologue.
We derived $T_{\rm {ex}}$ to be $6.2 \pm 0.6$, $6.2 \pm 0.6$, $6.3 \pm 0.5$, $6.3 \pm 0.6$, and $6.3 \pm 0.5$ K (1$\sigma$) for H$^{13}$CCCCCN, HC$^{13}$CCCCN, HCC$^{13}$CCCN, HCCC$^{13}$CCN, and HCCCC$^{13}$CN, respectively, from the intensities of the observed two rotational transitions.
These results are consistent with the value obtained from the normal species (6.5 K) within the errors, which implies that we can neglect the effect of self-trapping for the normal species.
We then calculated the column densities using $T_{\rm {ex}}$ obtained for each $^{13}$C isotopologue.
The $^{13}$C isotopic fractionation is not affected by the changes in $T_{\rm {ex}}$, and the $^{12}$C/$^{13}$C ratios are also not affected within the errors.
The results are also shown in Table \ref{tab2}.






\section{Discussion}

\subsection{Formation mechanisms of HC$_{5}$N}

\begin{figure}
\plotone{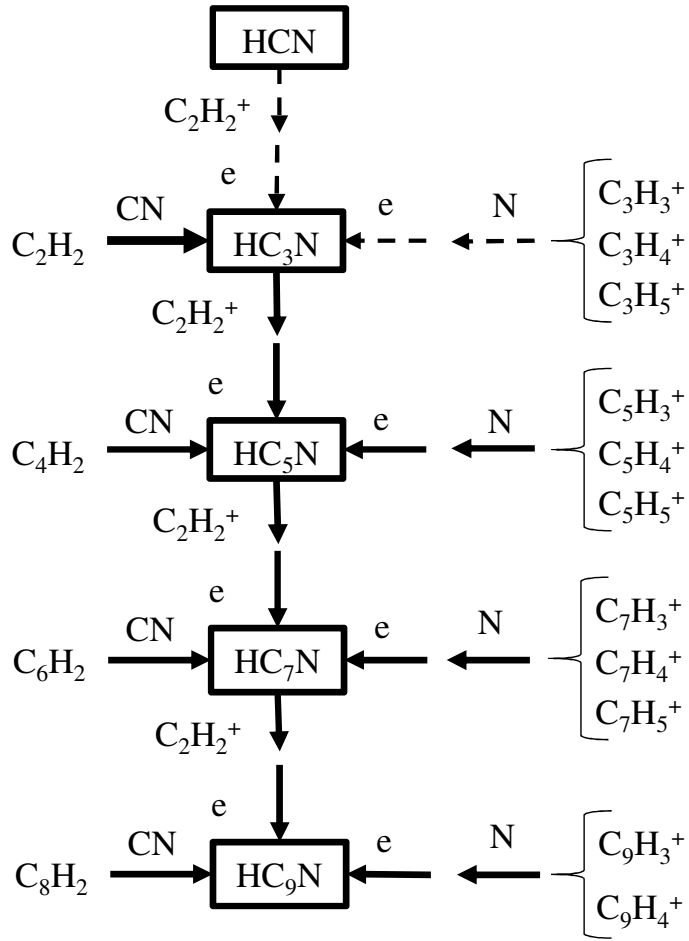}
\caption{Formation pathways leading to cyanopolyynes. The heavy line and dotted lines show the main and less contributed pathways, respectively \citep{tak98}.}  
\label{fig3}
\end{figure}

We here consider possible reactions leading to cyanopolyynes, using the UMIST Database for Astrochemistry 2012 \citep{mce13}.
By focusing on the rate constants, activation energies, abundances and formation pathways of precursors, we can sketch formation pathways of cyanopolyynes, as shown in Figure \ref{fig3}.  
Three types of formation pathways are possible as follows:
\\Pathway 1: the reactions of C$_{2n}$H$_{2}$ + CN,
\\Pathway 2: the growth of the cyanopolyyne carbon chains via C$_{2}$H$_{2}$$^{+}$ + HC$_{2n+1}$N, and
\\Pathway 3: the reactions between hydrocarbon ions and nitrogen atoms followed by electron recombination reactions. 

For HC$_{3}$N, Pathway 1 is suggested to play a dominant role in its production: \citet{tak98} reported that the main formation pathway of HC$_{3}$N is the reaction of C$_{2}$H$_{2}$ + CN on the basis of their observational results of $^{13}$C isotopic fractionation.
Fractionation of $^{13}$C in CN proceeds through the following reaction \citep{kai91}:
\begin{equation} \label{CN}
^{13}{\mathrm C}^{+} + {\mathrm {CN}} \rightarrow {\mathrm C}^{+} + ^{13} \! {\mathrm {CN}} + \Delta E \: (34 \: {\mathrm K}). 
\end{equation}  
This reaction is exothermic with energy of 34 K due to decreasing the zero-point vibrational energy.
Hence, its backward reaction is not effective in cold dark clouds, where the typical temperature is approximately 10 K \citep[e.g.,][]{ben89}. 
As a result, the $^{12}$C/$^{13}$C ratio for CN decreases in comparison with those for hydrocarbons such as C$_{2}$H$_{2}$ and C$_{4}$H$_{2}$, because similar fractionation processes of $^{13}$C do not occur in hydrocarbons \citep{wat76}.

If HC$_{5}$N is mainly formed by Pathways 1 and/or 2, at least one carbon atom in HC$_{5}$N, possibly that adjacent to the nitrogen atom, is not equivalent to the others. 
Assuming that HC$_{5}$N is mainly formed by Pathway 2, the $^{13}$C isotopic fractionation of HC$_{5}$N should be $x:y:1.0:1.0:1.4$, where $x$ and $y$ are arbitrary values, for $[$H$^{13}$CCCCCN$]: [$HC$^{13}$CCCCN$]: [$HCC$^{13}$CCCN$]: [$HCCC$^{13}$CCN$]: [$HCCCC$^{13}$CN$]$, unless scrambling of the carbon atoms is efficient.
However, our observed abundances are roughly equal for all the $^{13}$C isotopologues.
It is not likely that Pathway 1 and 2 are the main formation pathways of HC$_{5}$N.

On the basis of the above consideration, we propose that  Pathway 3 should significantly contribute to formation of HC$_{5}$N.
In Pathway 3, all carbon atoms in HC$_{5}$N originate from the hydrocarbon ions.
If all the carbon atoms in the mother hydrocarbon ion have the similar $^{12}$C/$^{13}$C ratio, the observational result can be explained.
This seems indeed possible for a large hydrocarbon ion, because it is generally produced through various processes.
This pathway is similar to the chemical model calculation by \citet{mar00}, where HC$_{5}$N can be produced through the reaction between a hydrocarbon ion (C$_{5}$H$_{3}$$^{+}$) and a nitrogen atom.
Thus, we propose that Pathway 3 overwhelms Pathways 1 and 2, resulting in almost equivalent abundances of the $^{13}$C isotopologues of HC$_{5}$N.

Based on the above discussion on HC$_{5}$N, we also propose that the longer cyanopolyynes may be formed by a mechanism similar to that of HC$_{5}$N, namely the reactions between hydrocarbon ions and nitrogen atoms.
However, it is difficult to confirm our suggestion by observations of $^{13}$C isotopic fractionation at the present.
\citet{lan07} tentatively detected the $^{13}$C isotopologues of HC$_{7}$N using the {\it J} = $11-10$ and $12-11$ transitions for the first time in the interstellar medium.
They calculated the averaged $^{12}$C/$^{13}$C ratio of HC$_{7}$N, but they could not study its $^{13}$C isotopic fractionation as they only derived upper limits for each of the seven $^{13}$C isotopologues.
Hence, more sensitive observations are needed to confirm the flat $^{13}$C abundance over the seven $^{13}$C isotopologues of HC$_{7}$N.
In the theoretical model by \citet{mar00}, the pathways leading to HC$_{7}$N and HC$_{9}$N via reactions between hydrocarbon ions (C$_{7}$H$_{3}$$^{+}$ and C$_{9}$H$_{3}$$^{+}$, respectively) and nitrogen atoms followed by the electron recombination reactions are presented.
The most interesting point to emerge from our analysis is that cyanopolyynes longer than HC$_{5}$N are mainly produced by the reactions between hydrocarbon ions and nitrogen atoms, whereas HC$_{3}$N is mainly produced by the neutral-neutral reaction between C$_{2}$H$_{2}$ and CN.
One possible reason is high abundance of C$_{2}$H$_{2}$ \citep{mar00}.

\subsection{Dilution of the $^{13}$C species}\label{dil}

We compile $^{12}$C/$^{13}$C ratios of some carbon-chain molecules at TMC-1 CP in Table \ref{tab3}.
The $^{12}$C/$^{13}$C ratios of carbon-chain molecules show more or less higher than that of the local interstellar elemental ratio ($60-70$) \citep[e.g.,][]{lan90,lan93,sav02,mil05}.
This can be explained by the following reaction, which should be effective under low-temperature conditions \citep[e.g.,][]{wat76,lan84},
\begin{equation} \label{CO}
^{13}{\mathrm C}^{+} + {\mathrm {CO}} \rightarrow {\mathrm C}^{+} + ^{13}\!{\mathrm {CO}} + \Delta E \: (35 \: {\mathrm K}). 
\end{equation}   
The reaction leads to a reduction in $^{13}$C$^{+}$.
Consequently, $^{13}$C isotopologues of carbon-chain molecules are also reduced, because C$^{+}$ and C are important raw materials of carbon-chain molecules.

We derived the five $^{12}$C/$^{13}$C ratios of HC$_{5}$N, and they are slightly higher than those of HC$_{3}$N (see Table \ref{tab3}).
\citet{lan07} were able to detect the $^{13}$C isotopologues only by combining the observations of all seven isotopologues of HC$_{7}$N into one spectrum.
They derived the averaged $^{12}$C/$^{13}$C ratio of HC$_{7}$N by combination of multiple lines of the seven $^{13}$C isotopologues of HC$_{7}$N.
These ratios indicate that the $^{13}$C species of the cyanopolyynes are less diluted than other carbon-chain molecules.
For example, for HC$_{3}$N and HC$_{5}$N, the $^{12}$C/$^{13}$C ratios coincide among their $^{13}$C isotopologues within about 20\% and 10\%, respectively.
On the other hand, for CCS, $^{12}$C/$^{13}$C of $^{13}$CCS is about 4 times larger than that of C$^{13}$CS.
In other words, an interesting characteristic of the $^{12}$C/$^{13}$C ratios of the two cyanopolyynes is that their differences appear to be small among the $^{13}$C isotopologues of each cyanopolyyne, compared to other carbon-chain molecules. 
The reason why the $^{13}$C species of the cyanopolyynes are not significantly diluted is unclear.
Mechanisms of the dilution of the $^{13}$C species are still controversial and are left for future studies.




\section{Conclusions}

We carried out observations of the {\it J} = $9-8$ and $16-15$ rotational transitions of the normal species and the five $^{13}$C isotopologues of HC$_{5}$N toward TMC-1 CP with the NRO 45-m telescope.
The abundance ratios of the five $^{13}$C isotopologues are found to be $1.00:0.97:1.03:1.05:1.16$ ($\pm 0.19$) (1$\sigma$) for $[$H$^{13}$CCCCCN$]: [$HC$^{13}$CCCCN$]: [$HCC$^{13}$CCCN$]: [$HCCC$^{13}$CCN$]: [$HCCCC$^{13}$CN$]$.
Based on the result, we suggest that reactions between hydrocarbon ions and nitrogen atoms play an important role in the formation of HC$_{5}$N.
We derived the $^{12}$C/$^{13}$C ratios from the five $^{13}$C isotopologues of HC$_{5}$N.
The averaged value is $94 \pm 6$ (1$\sigma$), and the $^{13}$C species of cyanopolyynes are less diluted than other carbon-chain molecules.

\acknowledgments

We are grateful to the staff of Nobeyama Radio Observatory.
Nobeyama Radio Observatory is a branch of the National Astronomical Observatory of Japan, National Institutes of Natural Sciences.
In particular, we deeply appreciate Mr. Jun Maekawa for developing a software for smoothed bandpass calibration analysis, and Dr. Izumi Mizuno for discussing the method of smoothed bandpass calibration analysis.
We express our thanks to Prof. Kazuhito Dobashi and Dr. Tomomi Shimoikura (Tokyo Gakugei University) for discussing  analysis of the data obtained by using the Z45 receiver, and to all the Z45 receiver group members for their kind support.
The Z45 receiver is supported in part by a Grant-in-Aid for Science Research of Japan (24244017) and the National Science Foundation under Grant (NSF PHY11-25915).

\clearpage



\clearpage

\clearpage

\clearpage

\begin{deluxetable}{lccccccc}
\tabletypesize{\scriptsize}
\tablecaption{Spectral line parameters of HC$_{5}$N and its five $^{13}$C isotopologues\label{tab1}}
\tablewidth{0pt}
\tablehead{
\colhead{Species} & \colhead{Transition} & \colhead{Frequency\tablenotemark{a}} & \colhead{{\it T}$^{\ast}_{\mathrm A}$\tablenotemark{b}} & \colhead{$\Delta v$} & \colhead{{\it V}$_{\mathrm {LSR}}$} & \colhead{$\int T^{\ast}_{\mathrm A}dv$\tablenotemark{c}} &  \colhead{rms\tablenotemark{d}} 
\\
\colhead{} & \colhead{} & \colhead{[GHz]} & \colhead{[mK]} & \colhead{[km s$^{-1}$]} & \colhead{[km s$^{-1}$]} & \colhead{[K km s$^{-1}$]} & \colhead{[mK]}
}
\startdata
HC$_{5}$N & $9-8$ & 23.9639007 (1) & 1600 (20) & 0.88 (1) & 5.90 (9) & 1.50 (3) & 2.6 \\ 
& $16-15$ & 42.6021529 (2) & 1874 (57) & 0.46 (2) & 5.9 (2) & 0.93 (4) & 3.3 \\
H$^{13}$CCCCCN & $9-8$ & 23.3400887 (5) & 44 (2) & 0.57 (3) & 6.0 (3) & 0.027 (2) & 2.2 \\
& $16-15$ & 41.4931720 (9) & 31 (3) & 0.44 (4) & 5.8 (5) & 0.0147 (19) & 3.2 \\
HC$^{13}$CCCCN & $9-8$ & 23.7183241 (5) & 35 (2) & 0.67 (4) & 5.8 (3) & 0.025 (2) & 2.4 \\
& $16-15$ & 42.1655778 (9) & 31 (3) & 0.44 (4) & 5.8 (4) & 0.0148 (16) & 3.0 \\ 
HCC$^{13}$CCCN & $9-8$ & 23.9390819 (6) & 25 (1) & 0.97 (6) & 5.8 (3) & 0.026 (2) & 2.1 \\
& $16-15$ & 42.5580316 (9) & 31 (3) & 0.49 (4) & 5.8 (4) & 0.0165 (16) & 3.1 \\
HCCC$^{13}$CCN & $9-8$ & 23.9419974 (5) & 31 (2) & 0.81 (5) & 5.8 (3) & 0.027 (2) & 2.1 \\
& $16-15$ & 42.5632150 (9) & 35 (3) & 0.45 (4) & 5.9 (4) & 0.0166 (19) & 3.1 \\
HCCCC$^{13}$CN & $9-8$ & 23.7271659 (6) & 43 (2) & 0.69 (4) & 5.9 (3) & 0.031 (2) & 2.4 \\
& $16-15$ & 42.1812962 (10) & 35 (3) & 0.47 (4) & 5.8 (5) & 0.0172 (18) & 3.0 \\
\enddata
\tablecomments{The numbers in parenthesis represent one standard deviation in the Gaussian fit except for frequency.}
\tablenotetext{a}{Taken from the Cologne Database for Molecular Spectroscopy (CDMS) \citep{mul05}.}
\tablenotetext{b}{We calibrated the peak intensities of the {\it J} = $16-15$ rotational transitions. We divided the results of Gaussian fit by 1.3, which was obtained by comparing \citet{suz92} and \citet{nak15}.}
\tablenotetext{c}{We calculated the integrated intensities and their errors using {\it T}$^{\ast}_{\mathrm A}$ and $\Delta v$, which are determined by the Gaussian fit, where the errors are estimated from those of {\it T}$^{\ast}_{\mathrm A}$ and $\Delta v$.}
\tablenotetext{d}{The rms noises in emission-free regions.}
\end{deluxetable}


\clearpage
\begin{deluxetable}{lcccc}
\tabletypesize{\scriptsize}
\tablecaption{Column densities and $^{12}$C/$^{13}$C ratios of HC$_{5}$N\label{tab2}}
\tablewidth{0pt}
\tablehead{
\colhead{Species} & \colhead{Column Density\tablenotemark{a}} & \colhead{$^{12}$C/$^{13}$C\tablenotemark{a}} & \colhead{Column Density\tablenotemark{b}} & \colhead{$^{12}$C/$^{13}$C\tablenotemark{b}}
\\
\colhead{} & \colhead{[$\times 10^{11}$ cm$^{-2}$]} & \colhead{ratio} & \colhead{[$\times 10^{11}$ cm$^{-2}$]} & \colhead{ratio}
}
\startdata
HC$_{5}$N & (6.2 $\pm$ 0.3) $\times 10^{2}$ & & & \\ 
H$^{13}$CCCCCN & $6.4 \pm 0.9$ & $98 \pm 14$ & $7.0 \pm 1.0$ & $89 \pm 13$ \\
HC$^{13}$CCCCN & $6.2 \pm 0.8$ & $101 \pm 14$ & $6.8 \pm 0.9$ & $91 \pm 12$ \\
HCC$^{13}$CCCN & $6.6 \pm 0.8$ & $95 \pm 12$ & $6.8 \pm 0.9$ & $91 \pm 12$ \\
HCCC$^{13}$CCN & $6.7 \pm 0.9$ & $93 \pm 13$ & $6.9 \pm 1.0$ & $90 \pm 13$ \\
HCCCC$^{13}$CN & $7.4 \pm 0.9$ & $85 \pm 11$ & $7.5 \pm 1.0$ & $83 \pm 11$ \\
\enddata
\tablecomments{The error corresponds to one standard deviation.}
\tablenotetext{a}{The values of $^{13}$C isotopologues were calculated by using $T_{\rm {ex}} = 6.5$ K obtained from the normal species.}
\tablenotetext{b}{The values were calculated by using the $T_{\rm {ex}}$ derived for each isotopologue.}  
\end{deluxetable}



\clearpage
\begin{deluxetable}{lcc}
\tabletypesize{\scriptsize}
\tablecaption{The $^{12}$C/$^{13}$C ratios in various carbon-chain molecules toward TMC-1 CP \label{tab3}}
\tablewidth{0pt}
\tablehead{
\colhead{Species} & \colhead{$^{12}$C/$^{13}$C ratio} & \colhead{Reference}
}
\startdata
HC$_{3}$N/H$^{13}$CCCN & 79 $\pm$ 11 (1$\sigma$) & 1 \\
HC$_{3}$N/HC$^{13}$CCN & 75 $\pm$ 10 (1$\sigma$) & 1 \\
HC$_{3}$N/HCC$^{13}$CN & 55 $\pm$ 7 (1$\sigma$) & 1 \\
HC$_{5}$N/H$^{13}$CCCCCN & 98 $\pm$ 14 (1$\sigma$) & 2 \\
HC$_{5}$N/HC$^{13}$CCCCN & 101 $\pm$ 14 (1$\sigma$) & 2 \\
HC$_{5}$N/HCC$^{13}$CCCN & 95 $\pm$ 12 (1$\sigma$) & 2 \\
HC$_{5}$N/HCCC$^{13}$CCN & 93 $\pm$ 13 (1$\sigma$) & 2 \\
HC$_{5}$N/HCCCC$^{13}$CN & 85 $\pm$ 11 (1$\sigma$) & 2 \\
HC$_{7}$N/average $^{13}$C isotopologues & 87 $^{+35}_{-19}$ (1$\sigma$) & 3 \\ 
CCH/$^{13}$CCH & \textgreater  250 & 4 \\
CCH/C$^{13}$CH & \textgreater 170 & 4 \\
C$_{4}$H/$^{13}$CCCCH & 141 $\pm$ 44 (3$\sigma$) & 5 \\
C$_{4}$H/C$^{13}$CCCH & 97 $\pm$ 27 (3$\sigma$) & 5 \\
C$_{4}$H/CC$^{13}$CCH & 82 $\pm$ 15 (3$\sigma$) & 5 \\
C$_{4}$H/CCC$^{13}$CH & 118 $\pm$ 23 (3$\sigma$) & 5 \\
CCS/$^{13}$CCS & 230 $\pm$ 130 (3$\sigma$) & 6 \\
CCS/C$^{13}$CS & 54 $\pm$ 5 (3$\sigma$) & 6 \\
C$_{3}$S/$^{13}$CCCS & \textgreater 206 (3$\sigma$) & 5 \\
C$_{3}$S/C$^{13}$CCS & 48 $\pm$ 15 (3$\sigma$) & 5 \\
C$_{3}$S/CC$^{13}$CS & 30 - 206 & 5 \\
\enddata
\tablerefs{
(1) Takano et al. 1998; (2) This work (3) Langston \& Turner 2007;
(4) Sakai et al. 2010; (5) Sakai et al. 2013; (6) Sakai et al. 2007.
}
\end{deluxetable}




\end{document}